# A case for polar uranium octupoles in cubic $U_2N_3$

S. W. Lovesey [1,2,3]

[1]*ISIS Facility, STFC, Didcot, Oxfordshire OX11 0QX, United Kingdom*
[2]*Diamond Light Source, Harwell Science and Innovation Campus, Didcot, Oxfordshire OX11 0DE, United Kingdom*
[3] *Department of Physics, Oxford University, Oxford OX1 3PU, UK*

**Abstract** Uranium ions in the sesquinitride alpha-$U_2N_3$ occupy independent acentric and centrosymmetric sites according to conventional x-ray diffraction patterns [R. Troć, J. Solid State Chem. **13**, 14 (1975)]. We submit that polar uranium multipoles in acentric sites are revealed in resonant x-ray diffraction data recently published by Lawrence Bright *et al*. [Phys. Rev. B **100**, 134426 (2019)]. To this end, their diffraction data gathered with a primary x-ray energy in the vicinity of the uranium $M_4$ absorption edge are compared to symmetry-informed diffraction amplitudes calculated for the bixbyite alpha-$Mn_2O_3$ lattice structure. Bragg spots forbidden in this lattice diffraction pattern appear to provide clear-cut evidence for high-order polar uranium multipoles.

## I. INTRODUCTION

X-ray diffraction patterns gathered on crystalline materials can contain Bragg spots that do not exist in patterns created by spheres of atomic charge located at points on the particular lattice. Their inherent weakness is off-set by tuning the energy of primary x-rays from a synchrotron source to a specific atomic resonance. The process is often called T & T scattering to acknowledge early experimental studies by Templeton and Templeton [1-5], or anisotropic tensor scattering [6, 7, 8]. The weak Bragg spots are not indexed by Miller indices for the lattice symmetry, i.e., they are space-group forbidden. Departures from spheres of atomic charge are usually labelled by components of an electronic quadrupole (multipole rank = 2) that are invariant with respect to operations in the symmetry of sites occupied by the resonant ions (Neumann's Principle [9]). Specifically, acentric sites in a crystal can harbour polar (parity-odd) multipoles. In our study of space-group forbidden reflections by (cubic) α-$U_2N_3$ we encounter uranium dipoles (rank = 1), quadrupoles and octupoles (rank = 3) that must be present in a meaningful model of the electronic structure.

Following Lawrence Bright *et al*., we use the bixbyite α-$Mn_2O_3$ type-lattice depicted in Fig. 1 with uranium ions in acentric sites 24d (U2) and centrosymmetric sites 8b (U1) [10, 11, 12]. The authors mention forbidden reflections (1, 1, 0), (3, 1, 0) and (3, 3, 0), and display data for the (0, 1, 3) Bragg spot, which is equivalent to (3, 1, 0) in cubic bixbyite.

## II. LATTICE STRUCTURE

Space group Ia$\bar{3}$ (No. 206) describes the (cubic) bixbyite alpha-Mn$_2$O$_3$ type-lattice [10, 13]. A body centre translation requires that integer Miller indices *h*, *k*, *l* have an even sum. Equivalent reflections for No. 206 are listed in Appendix A. A magnetic diffraction pattern observed below T$_N$ ≈ 73.5 K is consistent with an anti-translation and odd ($h + k + l$) [12]. The measured lattice constant $a$ ≈ 10.69 Å.

Lawrence Bright *et al*. [12] made measurements on thin epitaxial films of U$_2$N$_3$, and report in-plane lattice parameters that are different to the growth directions. The implication is that for their thin films, where the mosaicity is low, the symmetry is metrically orthorhombic and not cubic. For thicker films where the mosaicity is larger, this might mean the induced orthorhombic strain has been relaxed and the bulk of a film is cubic. Diffraction by an appropriate orthorhombic lattice symmetry appears in Section V, and we continue to study diffraction by uranium ions in cubic Ia$\bar{3}$ (No. 206).

## III. RESONANT X-RAY DIFFRACTION

States of x-ray polarization, Bragg angle θ, and the plane of scattering are shown in Fig. 2. The x-ray scattering length in the unrotated channel of polarization σ → σ', say, is modelled by (σ'σ)/D(E). In this instance, the resonant denominator is replaced by a sharp oscillator D(E) = {[E − Δ + $i$Γ/2]/Δ} with the x-ray energy E in the near vicinity of an atomic resonance Δ of total width Γ, namely, E ≈ Δ and Γ << Δ [14]. The cited energy-integrated scattering amplitude (σ'σ), one of four amplitudes, is studied using standard tools and methods from atomic physics and crystallography. Resonant processes from M-edges utilize atomic states 3d - 5f (electric dipole, E1) and 3d - 6s, 6d (electric quadrupole, E2). The meaningful comparison of E1 and E2 radial integrals in diffraction amplitudes includes the photon wavelength λ in E2 [14]. In the present case, (3d|R|5f) should be compared with [(2π/λ) (3d|R$^2$|6s)] or [(2π/λ) (3d|R$^2$|6d)]. One finds the dimensionally correct E2 integrals are a factor ≈ 16 smaller than the standard E1 radial integral (3d|R|5f) for λ ≈ 3.33 Å (based on Δ ≈ 3.73 keV). An E1-E1 (E1-E2) scattering event is parity-even (parity-odd) and exposes axial (polar) electronic properties.

In our adopted description of electronic degrees of freedom, uranium ions are assigned spherical multipoles ⟨O$^K_Q$⟩ of integer rank K with projections Q in the interval −K ≤ Q ≤ K. A unit-cell electronic structure factor Ψ$^K_Q$ is compiled from all symmetry operations for uranium ions in space group No. 206 [13]. Our results for Ψ$^K_Q$ (24d) and Ψ$^K_Q$ (8b) are reported in Appendix A. The main text of our paper is given over to a discussion of key implications for weak (space- group forbidden) Bragg spots available from resonant x-ray diffraction.

Cartesian and spherical components Q = 0, ±1 of a vector **n** = (ξ, η, ζ) are related by ξ = (n$_{-1}$ − n$_{+1}$)/√2, η = $i$(n$_{-1}$ + n$_{+1}$)/√2, ζ = n$_0$. A complex conjugate of a multipole is defined as ⟨O$^K_Q$⟩* = (−1)$^Q$ ⟨O$^K_{-Q}$⟩, meaning the diagonal multipole ⟨O$^K_0$⟩ is purely real. The phase convention for real and imaginary parts labelled by single and double primes is $\langle O^K_Q \rangle = [\langle O^K_Q \rangle'$ $+ i\langle O^K_Q \rangle'']$. Whereupon, Cartesian dipoles are $\langle O^1_\xi \rangle = -\sqrt{2} \langle O^1_{+1} \rangle'$ and $\langle O^1_\eta \rangle = -\sqrt{2} \langle O^1_{+1} \rangle''$.

## IV. WEAK REFLECTIONS: Cubic $Ia\bar{3}$

We consider a Bragg reflection vector $(h, k, 0)$ with even $(h + k)$ in keeping with Table I in Ref. [12]. Odd integer Miller indices $h$ and $k$ produce weak reflections, e.g., $\Psi^K_Q(8b) = 0$ for the projection $Q = 0$, and $(1, 1, 0)$, $(3, 1, 0)$ and $(3, 3, 0)$ are specifically mentioned [12]. The electronic structure factor Eq. (A2) for $\Psi^K_Q(8b)$ can different from zero for odd Q, however. Axial multipoles have an even rank and $K = 2$ in an E1-E1 absorption event [15]. Since the quadrupoles $\langle T^2_{\pm 1} \rangle_b$ are not invariant with respect to $3^\pm_{xyz}$ in the full 8b site symmetry uranium ions in the centrosymmetric sites do not contribute to the defined pattern of weak Bragg spots. We continue the discussion of weak Bragg spots with diffraction by uranium ions in acentric 24d sites.

Evaluated for $(h, k, 0)$ the electronic structure factor Eq. (A3) for $\Psi^K_Q(24d)$ reduces to,

$$\Psi^K_Q(24d) = \langle O^K_Q \rangle_d \, [\alpha(h) + \alpha(h)^* \, \sigma_\pi] \, [1 - (-1)^Q \, \sigma_\pi] \qquad (2n+1, 2m+1, 0)$$

$$+ 2 \, \alpha(k)' \, \gamma(h) \, 3^+_{xyz} \langle O^K_Q \rangle_d \, [1 - \sigma_\pi]. \qquad (1)$$

Spatial phase factors are $\alpha(h) = \exp(i2\pi hx)$, $\gamma(h) = \exp(i\pi h/2)$. Notably, multipoles $3^-_{xyz} \langle O^K_Q \rangle_d$ are wholly absent in Eq. (1), and axial $3^+_{xyz} \langle T^K_Q \rangle_d$ with parity signature $\sigma_\pi = +1$ are absent.

Universal results for diffraction amplitudes are listed in Ref. [15] as functions of two quantities even and odd with respect to the sign of Q, namely, $A^K_{-Q} = A^K_Q$ and $B^K_{-Q} = -B^K_Q$. They are functions of the azimuthal angle $\psi$ (angle of rotation of the crystal about the reflection vector), with the crystal c axis normal to the plane of scattering at the beginning of a scan $\psi = 0$. For the case in hand (equivalent reflections for space group $Ia\bar{3}$ (No. 206) are listed in the Appendix),

$$A^K_Q + B^K_Q = e^{(iQ\chi)} \Psi^K_Q, \qquad (h, k, 0) \qquad (2)$$

where $\cos(\chi) = -h/\sqrt{[h^2 + k^2]}$.

Axial multipoles contribute standard T & T scattering proportional to the uranium quadrupole $\langle T^2_Q \rangle_d$ with odd Q [1-5]. The E1-E1 amplitude $(\sigma'\sigma)_{11}$ is purely real,

$$(\sigma'\sigma)_{11} = \cos(\chi) \sin(2\psi) \, \alpha(h)' \, \langle T^2_{+1} \rangle_d'', \qquad (h, k, 0) \qquad (3)$$

with $\alpha(h)' = \cos(2\pi hx)$. The diffraction amplitude in which x-ray polarization is rotated from $\sigma$ to $\pi'$ contains an additional contribution [15],

$$(\pi'\sigma)_{11} = \alpha(h)' \langle T^2_{+1}\rangle_d'' [\sin(\theta) \cos(\chi) \cos(2\psi) + \cos(\theta) \sin(\chi) \cos(\psi)], \quad (h, k, 0) \quad (4)$$

and $(\sigma'\pi)_{11}$ is obtained by changing the sign of the Bragg angle $\theta$.

Uranium polar multipoles are denoted $\langle U^K_Q\rangle$, and we henceforth safely drop the subscript d given that they are exclusive to 24d. It is convenient to write the E1-E2 amplitude $(\sigma'\sigma)_{12}$ as a sum of two parts, one proportional to $\alpha(h)''$ (labelled V) and one proportional to $\alpha_k(k)'$ (labelled W), i.e., $(\sigma'\sigma)_{12} = V + [\gamma(h) W]$. The amplitude $(\sigma'\sigma)_{12}$ depends on the Bragg angle unlike $(\sigma'\sigma)_{11}$. We find V is a sum of multipoles with even projections, two quadrupoles (rank K = 2) and an octupole (rank K = 3) [15],

$$V = (1/\sqrt{30}) \alpha(h)'' \sin(\theta) \sin(2\psi) [\sqrt{(3/2)} \langle U^2_0\rangle$$

$$+ \cos(2\chi) \{\langle U^2_{+2}\rangle' + 2\sqrt{2} \langle U^3_{+2}\rangle''\}]. \quad (h, k, 0) \quad (5)$$

There are seven purely imaginary contributions to W, which is proportional to $\sin(\theta)$. The factor $\gamma(h) = \exp(i\pi h/2)$ renders $[\gamma(h) W]$ in $(\sigma'\sigma)_{12}$ real for odd $h$. An actual expression for W can be read off from Eq. (D1) in Ref. [15] by inserting multipoles listed in an Appendix to this paper. For the special case $\cos(2\chi) = 0$ and $h = 2n +1$,

$$(\sigma'\sigma)_{12} = \sin(\theta) [(1/\sqrt{20}) \alpha(h)'' \sin(2\psi) \langle U^2_0\rangle \quad (2n+1, 2n+1, 0)$$

$$- (i/5) \sqrt{6} \gamma(h) \alpha(h)' \{\hat{E}^1_1 + (1/2) \sqrt{(5/6)} \sin(2\psi) \hat{A}^2_0$$

$$+ (i/3) \sqrt{5} \cos(2\psi) \hat{A}^2_1 + (1/6) [5 \cos(2\psi) + 3] \hat{E}^3_1 - \sqrt{(5/3)} \sin^2(\psi) \hat{E}^3_3\}]. \quad (6)$$

Polar dipoles, quadrupoles and octupoles $\hat{E}^1_1$, $\hat{A}^2_Q$, $\hat{E}^3_Q$ appearing in Eq. (6) are listed in Eq. (A6). In summary, the dependence of $(\sigma'\sigma)_{12}$ on $\psi$ is a linear combination of $\cos(2\psi)$ and $\sin(2\psi)$ on a background set by a dipole $\hat{E}^1_1$ directed along the a axis in Fig. 1.

### V. REDUCED LATTICE SYMMETRY: Ibca

The parent cubic space group $Ia\bar{3}$ (No. 206) does not contain any fourfold axes. In consequence, a strained film can only be orthorhombic. If we take into account the orthorhombic macroscopic strains (imposed by the misfit with the substrate) the reduced symmetry is orthorhombic Ibca (No. 73) that keeps the lattice vectors and origin of the parent structure and origin. This lattice symmetry still forbids (3, 1, 0) or any other reflection $(h, k, 0)$ with odd $h, k$. Electronic structure factors for uranium ions are given in Appendix B. An accompanying list of equivalent reflections confirms that (3, 1, 0) and (1, 3, 0) are not equivalent in the orthorhombic space group. To begin with, we compare some forbidden $(h, k, 0)$ scattering amplitudes with foregoing results derived for the cubic space group $Ia\bar{3}$.

Uranium ions in sites 8b contribute diffraction enhanced by an E1-E1 event, whereas special sites do not contribute in the cubic lattice. Axial T & T scattering with odd $h$ and $k$ include,

$$(\sigma'\sigma)_{11} = \sin(\chi) \sin(2\psi) \langle T^2_{+1}\rangle_b', \qquad (2n+1, 2m+1, 0) \qquad (7)$$

$$(\pi'\sigma)_{11} = \langle T^2_{+1}\rangle_b' [\sin(\theta) \sin(\chi) \cos(2\psi) - \cos(\theta) \cos(\chi) \cos(\psi)]. \qquad (8)$$

The amplitude $(\sigma'\pi)_{11}$ is obtained from Eq. (8) by a change in sign of the Bragg angle $\theta$. Contributions to T & T scattering in $(\sigma'\sigma)_{11}$ from sites 8c and 8d are proportional to $\sin(2\psi)$, as in Eq. (7). The remaining factors in $(\sigma'\sigma)_{11}$ for the two sites are $[\alpha(h)' \cos(\chi) \langle T^2_{+1}\rangle_c'']$ and $[i\alpha(k)'' \gamma(h) \sin(\chi) \langle T^2_{+1}\rangle_d']$.

The polar amplitude $(\sigma'\sigma)_{12}$ for sites 8c is also proportional to $\sin(2\psi)$. An uranium octupole $\langle U^3_{+2}\rangle_c''$ is engaged,

$$(\sigma'\sigma)_{12} = (1/\sqrt{20}) \sin(\theta) \alpha(h)'' \sin(2\psi)$$

$$\times [\langle U^2_0\rangle_c + \cos(2\chi) \{\sqrt{(2/3)} \langle U^2_{+2}\rangle_c' + (4/\sqrt{3}) \langle U^3_{+2}\rangle_c''\}]. \quad (2n+1, 2m+1, 0) \quad (9)$$

The corresponding result for sites 8d is obtained by the substitution $\alpha(h)'' \rightarrow -i\alpha(k)' \gamma(h)$ along with $\langle U^K_Q\rangle_c \rightarrow \langle U^K_Q\rangle_d$. The octupole is a greater presence in the rotated channel for which we find,

$$(\pi'\sigma)_{12} = \text{polar quadrupoles} + (2/\sqrt{15}) \sin(\theta) \alpha(h)'' \langle U^3_{+2}\rangle_c''$$

$$\times [\cos(\theta) \sin(2\chi) \cos(\psi) + \sin(\theta) \cos(2\chi) \cos(2\psi)], \quad (2n+1, 2m+1, 0) \quad (10)$$

with a similar result for sites 8d.

Turning to site 8e, the electronic structure factor Eq. (B3) is zero for $\sigma\pi = +1$ using odd $k$ and $l = 0$. The polar amplitude $(\sigma'\sigma)_{12}$ can be different from zero, however, and an octupole is engaged. We find,

$$(\sigma'\sigma)_{12} = (i/\sqrt{30}) \sin(\theta) \alpha(h)'' \gamma(k) \sin(2\psi) \sin(2\chi)$$

$$\times [\langle U^2_{+2}\rangle_e'' - (4/\sqrt{3}) \langle U^3_{+2}\rangle_e'\}]. \qquad (2n+1, 2m+1, 0) \qquad (11)$$

There is not scope from the azimuthal angle dependences of $(\sigma'\sigma)_{12}$ to differentiated between sites. However, for site 8e, $(\sigma'\sigma)_{12} = 0$ for $\chi = 0$ while contributions are allowed from 8c and 8d for this reflection vector.

Scattering amplitudes for reflections $(0, k, l)$ with even $(k + l)$ and odd $k$ can be derived starting from Eq. (B4). Axial amplitudes for the special sites 8b are absent, namely, $(\sigma'\sigma)_{11} = (\pi'\sigma)_{11} = 0$. The electronic structure factor $\Psi^K_Q(8c)$ in Eq. (B1) vanishes for axial multipoles defined by $\sigma_\pi = +1$, while $(\sigma'\sigma)_{12} \propto \sin(2\chi)$ with $\cos(\chi) = k/\sqrt{[k^2 + l^2]}$. Axial T & T scattering by uranium ions in general sites 8d and 8e is performed by quadrupoles $\langle T^2_{+1}\rangle_d'$ and $\langle T^2_{+2}\rangle_e''$, respectively. Polar multipoles that arise in the unrotated polar channel $(\sigma'\sigma)_{12}$ are similar for the two sites. Specifically, $(\sigma'\sigma)_{12}$ contains a contribution from an octupole $\propto [\cos(2\chi) \langle U^3_{+2}\rangle'']$.

## VI. CONCLUSIONS

It is argued that the observation of a space-group forbidden Bragg spot in resonant x-ray diffraction by cubic $U_2N_3$ is solid evidence of uranium electronic multipoles in the material [8, 12, 17, 18]. Multipoles include a dipole parallel to the crystal a axis, and octupoles. Such results take our case beyond axial quadrupoles engaged in conventional Templeton - Templeton scattering, as in Eqs. (3) and (4) [1-5]. Looking ahead, our predicted azimuthal angle scans - intensity versus rotation of the sample about the reflection vector - need testing. The outcome could sway a debate about the relevant significance of axial and polar uranium multipoles in cubic $U_2N_3$, as we see in Eqs. (3) and (6) for diffraction amplitudes $(\sigma'\sigma)_{11}$ and $(\sigma'\sigma)_{12}$.

Electronic structure factors Eqs. (A2) and (A3) for uranium ions have been used mainly to interpret resonant x-ray diffraction in the unrotated channel of polarization $(\sigma'\sigma)$ [12]. Universal expressions for all four amplitudes in Ref. [15] permit an exhaustive investigation at the time data are available, using electric dipole - electric dipole (E1-E1), electric dipole - electric quadrupole (E1-E2) or E2-E2 absorption events. For the moment, we have used E1-E1 and E1-E2, and included a few statements about the rotated channel $(\pi'\sigma)$. Beyond resonant x-ray diffraction, structure factors Eqs. (A2) and (A3) can be used to calculate neutron diffraction patterns [16].

Our theoretical investigation is motivated by diffraction patterns collected from thin epitaxial films of $U_2N_3$ [12]. Lawrence Bright *et al*. report in-plane lattice parameters that are different to the growth directions. This piece of information plants the idea that the lattice symmetry for the diffraction patterns of interest is metrically orthorhombic and not cubic, an idea we pursue in Section V and Appendix B.

**Acknowledgements** I benefitted from correspondence with Dr K. S. Knight now reflected in Section II. Dr D. D. Khalyavin steered the study of reduced lattice symmetry and prepared Fig. 1. Radial integrals for x-ray diffraction quoted in Section III were calculated by Professor G. van der Laan using Cowan's atomic code [19, 20]. Dr G. H. Lander introduced me to work with his colleagues on cubic $U_2N_3$ [12].

# APPENDIX A: Cubic $Ia\bar{3}$ (No. 206)

An electronic structure factor,

$$\Psi^K_Q = [\exp(i\boldsymbol{\kappa}\cdot\mathbf{e})\langle O^K_Q\rangle_\mathbf{d}], \tag{A1}$$

where the reflection vector $\boldsymbol{\kappa} = (h, k, l)$ and the implied sum is over uranium ions in the unit cell at positions $\mathbf{e}$. Sites 8b in space group No. 206 (crystal class $m\bar{3}$) are centrosymmetric (Fig. 1) and axial multipoles are denoted by $\langle T^K_Q\rangle_b$ [14]. Additional site symmetries for 8b require $\langle T^K_Q\rangle_b$ to be invariant with respect to angular rotations $3^+_{xyz}$ and $3^-_{xyz}$. We find [13],

$$\Psi^K_Q(8b) = e^{\{i\pi(h+k+l)/2\}}[1 + (-1)^h(-1)^Q][\langle T^K_Q\rangle_b + (-1)^k(-1)^K\langle T^K_{-Q}\rangle_b]. \tag{A2}$$

Sites 24d possess symmetry $2_x$, and $2_x\langle O^K_Q\rangle_d = (-1)^K\langle O^K_{-Q}\rangle_d = (-1)^{K+Q}\langle O^K_Q\rangle_d^*$ [13]. Thus, a generic multipole $\langle O^K_Q\rangle_d$ is purely real (imaginary) for even (odd) K + Q. Spatial phase factors in $\Psi^K_Q(24d)$ include $\alpha(h) = \exp(i2\pi hx)$, $\gamma(h) = \exp(i\pi h/2)$, etc., with $x \approx -0.02$ [12]. A parity signature $\sigma_\pi = +1$ ($-1$) for axial (polar) multipoles. After some algebra,

$$\Psi^K_Q(24d) = \gamma(l)\langle O^K_Q\rangle_d[\alpha(h)+(-1)^l\alpha(h)^*\sigma_\pi][1+(-1)^h(-1)^Q\sigma_\pi]$$

$$+ \gamma(h)\,3^+_{xyz}\langle O^K_Q\rangle_d[\alpha(k)+(-1)^l\alpha(k)^*][1+(-1)^k\sigma_\pi]$$

$$+ \gamma(k)\,3^-_{xyz}\langle O^K_Q\rangle_d[\alpha(l)+(-1)^h\alpha(l)^*][1+(-1)^l\sigma_\pi]. \tag{A3}$$

A derivation of $\Psi^K_Q(24d)$ exploits the identities $3^+_{xyz}(x, y, z) = (z, x, y) = 2_x 3^+_{-xy-z}(x, y, z)$, and $3^-_{xyz}(x, y, z) = (y, z, x) = 2_x 3^-_{-x-yz}(x, y, z)$ (Appendix 1, Ref. [21]). In the unrealistic case of identical uranium ions at the two independent sites, $[\Psi^K_0(8b) + \Psi^K_0(24d)] = 0$ for even K, $(h, h, 0)$, $\gamma(h) = -1$ and $\sigma_\pi = +1$. Multipoles $3^\pm_{xyz}\langle O^K_Q\rangle$ in Eq. (A3) can be constructed with,

$$3^+_{xyz}\langle O^K_Q\rangle = \exp(iq\beta)\,d^K_{Qq}(\beta)\langle O^K_q\rangle, \tag{A4}$$

$$3^-_{xyz}\langle O^K_Q\rangle = \exp(iQ\beta)(-1)^q\,d^K_{Qq}(\beta)\langle O^K_q\rangle, \tag{A5}$$

A sum on projections q in Eqs. (A4) and (A5) is implied, and $d^K_{Qq}(\beta)$ is a standard Wigner rotation matrix with an argument $\beta = \pi/2$ [22]. The identity $d^K_{-Qq}(\beta) = (-1)^{K+q}d^K_{Qq}(\beta)$ is useful in calculations.

We list multipoles exposed at a reflection $(2n+1, 2m+1, 0)$ with $\cos(\chi) = -h/\sqrt{[h^2+k^2]}$. Polar uranium multipoles in W are defined by $(\sigma'\sigma)_{12} = [V + \gamma(h)W]$, and they are;

$$\hat{E}^1_1 = -\sin(\chi)\langle U^1_{+1}\rangle' = \sin(\chi)\langle U^1_a\rangle/\sqrt{2},$$

$$\hat{A}^2_2 = (1/2)\cos(2\chi)[\sqrt{(3/2)}\langle U^2_0\rangle - \langle U^2_{+2}\rangle'],\quad \hat{A}^2_1 = i\sin(\chi)\langle U^2_{+1}\rangle'',$$

$$\hat{A}^2{}_0 = -[(1/2)\langle U^2{}_0\rangle + \sqrt{(3/2)} \langle U^2{}_{+2}\rangle'],$$

$$\hat{E}^3{}_3 = (1/4)\sin(3\chi)[-\sqrt{15}\langle U^3{}_{+1}\rangle' + \langle U^3{}_{+3}\rangle'], \quad \hat{E}^3{}_2 = i\cos(2\chi)\langle U^3{}_{+2}\rangle'',$$

$$\hat{E}^3{}_1 = (1/4)\sin(\chi)[\langle U^3{}_{+1}\rangle' + \sqrt{15}\langle U^3{}_{+3}\rangle']. \tag{A6}$$

*Space group*: equivalent reflections for cubic No. 206;
(1) $h, k, l$; (2) $-h, -k, l$; (3) $-h, k, -l$; (4) $h, -k, -l$; (5) $l, h, k$; (6) $l, -h, -k$; (7) $-l, -h, k$;
(8) $-l, h, -k$; (9) $k, l, h$; (10) $-k, l, -h$; (11) $k, -l, -h$; (12) $-k, -l, h$; (13) $-h, -k, -l$; (14) $h, k, -l$;
(15) $h, -k, l$; (16) $-h, k, l$; (17) $-l, -h, -k$; (18) $-l, h, k$; (19) $l, h, -k$; (20) $l, -h, k$;
(21) $-k, -l, -h$; (22) $k, -l, h$; (23) $-k, l, h$; (24) $k, l, -h$. (A7)

## APPENDIX B: Orthorhombic Ibca (No. 73)

Miller indices $h, k, l$ have an even sum for space group Ibca (No. 73, crystal class mmm). Uranium ions occupy four independent sites, one special (8b) and three general (8c, 8d, 8e) [10, 13]. The electronic structure factor $\Psi^K{}_Q(8b)$ is identical to Eq. (A2) with site symmetry reduced from $\bar{3}$ in No. 206 to inversion alone in No. 73. The remaining three electronic structure factors are,

$$\Psi^K{}_Q(8c) = \gamma(l) \langle O^K{}_Q\rangle_c [\alpha(h) + (-1)^l \alpha(h)^* \sigma_\pi] [1 + (-1)^h (-1)^Q \sigma_\pi]. \tag{B1}$$

$$\Psi^K{}_Q(8d) = \gamma(h) \langle O^K{}_Q\rangle_d [\alpha(k) + (-1)^h \alpha(k)^* \sigma_\pi] [1 + (-1)^k (-1)^Q \sigma_\pi]. \tag{B2}$$

$$\Psi^K{}_Q(8e) = \gamma(k) \langle O^K{}_Q\rangle_c [\alpha(l) + (-1)^k \alpha(l)^* \sigma_\pi]$$

$$\times [\langle O^K{}_Q\rangle_e + (-1)^l (-1)^K \sigma_\pi \langle O^K{}_{-Q}\rangle_e]. \tag{B3}$$

Sites symmetries in Eqs. (B1), (B2) and (B3) are $2_x$ (8c), $2_y$ (8d) and $2_z$ (8e), leading to $\langle O^K{}_Q\rangle_c{}^* = (-1)^{K+Q} \langle O^K{}_Q\rangle_c$, $\langle O^K{}_Q\rangle_d{}^* = (-1)^K \langle O^K{}_Q\rangle_d$ and even Q (8e). The general coordinate in a spatial factor $\alpha(h) = \exp(i2\pi h y)$, say, is not known. Data for bixbyite $\alpha$-$Mn_2O_3$ suggests y ≈ −0.035 [10].

Scattering amplitudes for reflections $(0, k, l)$ with even $(k + l)$ and odd $k$ can be derived from,

$$A^K{}_Q + B^K{}_Q = \gamma(Q) e^{(iQ\chi)} d^K{}_{Qq}(\beta) \Psi^K{}_q, \quad (0, k, l) \tag{B4}$$

with a sum on projections q. In the present case, $\cos(\chi) = k/\sqrt{[k^2 + l^2]}$, $\beta = \pi/2$ and $\gamma(Q) = \exp(i\pi Q/2)$. The crystal a axis is parallel to $-z$ in Fig. 2 at the start of an azimuthal angle scan $\psi = 0$.

*Space group*: equivalent reflections for orthorhombic No. 73;

(1) *h*, *k*, *l*; (2) −*h*, −*k*, *l*; (3) −*h*, *k*, −*l*; (4) *h*, −*k*, −*l*;

(5) −*h*, −*k*, −*l*; (6) *h*, *k*, −*l*; (7) *h*, −*k*, *l*; (8) −*h*, *k*, *l*.

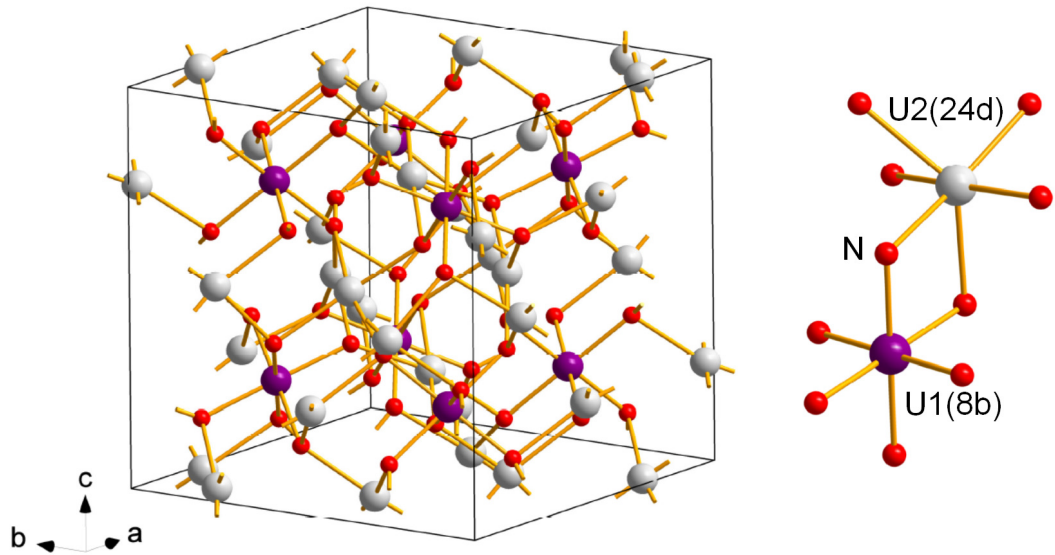

FIG. 1. The bixbyite structure No. 206 has six-fold coordinated cations (large spheres) occupying centrosymmetric 8b (U1) and acentric 24d (U2) Wyckoff sites [13]. Cation sites are surrounded by local octahedra with oxygen (small spheres) at their vertexes.

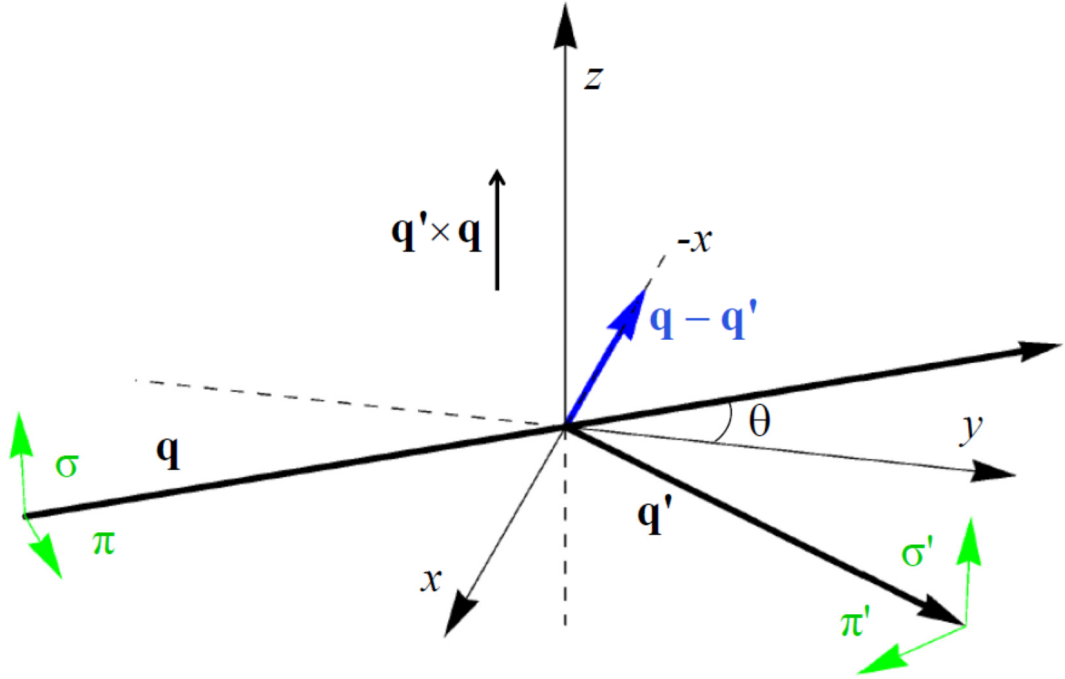

FIG. 2. Primary (σ, π) and secondary (σ', π') states of x-ray polarization. Corresponding wavevectors **q** and **q'** subtend an angle 2θ. The Bragg condition for diffraction is met when **q** − **q'** coincides with a reflection vector ($h, k, l$). Crystal vectors in Fig. 1 and the depicted Cartesian (x, y, z) coincide in the nominal setting of the crystal, and the start ψ = 0 of an azimuthal angle scan (rotation of the crystal by an angle ψ about the reflection vector).